\documentclass[preprint,aps,showpacs,nofootinbib,preprintnumbers,amsmath,amssymb]{revtex4-1}
\usepackage{epsfig}
\usepackage{bm}
\usepackage[usenames ,dvipsnames]{xcolor}
\usepackage{slashed}
\usepackage{graphicx,color}

\usepackage{multirow}
\begin{document}
	\title{Nonleptonic two-body weak decays of $\Lambda_b$ in modified MIT bag model}
	
	\author{Chao-Qiang Geng$^{1,2,3,4}$, Chia-Wei Liu$^{3}$, Tien-Hsueh Tsai$^{3}$ }
	\affiliation{
		$^{1}$School of Fundamental Physics and Mathematical Sciences, Hangzhou Institute for Advanced Study, UCAS, Hangzhou 310024, China \\
$^{2}$International Centre for Theoretical Physics Asia-Pacific, Beijing/Hangzhou, China \\
$^{3}$Department of Physics, National Tsing Hua University, Hsinchu 300, Taiwan\\
$^{4}$Physics Division, National Center for Theoretical Sciences, Hsinchu 300, Taiwan
}\date{\today}

	\begin{abstract}
We study the nonleptonic two-body weak decays of $\Lambda_b$ by modifying the MIT bag model without introducing new parameters to 
construct the momentum eigenstates of the baryons. We find that the branching ratios of $\Lambda_b^0 \to \Lambda_c^+ \pi^-$,  $\Lambda_c^+ K^-$, $p \pi^-$ and $pK^-$ are  $(4.5\pm0.2)\times 10^{-3}$, $(3.4\pm 0.1)\times 10^{-4}$, $(5.0\pm 0.5)\times 10^{-6}$ and $(6.0\pm 0.7) \times 10^{-6}$, which are all well consistent with the current experimental data, respectively. We also explore  P and CP  asymmetries for the decays of $\Lambda_b^0 \to p (\pi^-,K^-)$. In particular,  we obtain that the direct CP-violating rate asymmetries in $\Lambda_b^0 \to p \pi^-$ and $\Lambda_b^0 \to p K^-$ are 
around $-4.4\%$  and $6.7\%$,  
in comparison with $(-2.5\pm2.9)\%$ and $(-2.5\pm2.2)\%$ from the Particle Data Group in 2020,  respectively.
	\end{abstract}

\maketitle
\section{Introduction}
There have been many interesting measurements in the decays of $b$-baryons  by the LHCb Collaboration,
including those for the charmful modes of $\Lambda_b^0\to \Lambda_c^+ M$~\cite{Aaij:2014lpa} and
charmless ones of $\Lambda_b^0\to p M$~\cite{Aaltonen:2008hg} ($M=\pi^-,K^-$) as well as the discoveries of the hidden-charm pentaquarks
in $\Lambda_b^0\to J/\psi p M$~\cite{Aaij:2015fea,Aaij:2016ymb} and double-charm baryon state of $\Xi^{++}_{cc}$ via 
 $\Xi^{++}_{cc}\to \Lambda_c^+K^-\pi^+\pi^+$~\cite{LHCbDcharmB}.
In particular, based on the new experimental data from  LHCb~\cite{CPnew},
the Particle Data Group~(PDG) of 2020~\cite{pdg}
has updated the average  values for the direct CP-violating rate asymmetries (${\cal A}_{CP}$s)
in $\Lambda_b^0 \to p \pi^-$ and $\Lambda_b^0 \to p K^-$
 to be
\begin{eqnarray}
\label{ACPexpt}
{\cal A}_{CP} (\Lambda_b^0 \to p \pi^-)_{\rm PDG}=(-2.5\pm 2.9)\%\,,\nonumber\\
{\cal A}_{CP} (\Lambda_b^0 \to p K^-)_{\rm PDG} =(-2.5\pm 2.2)\%\,,
\end{eqnarray}
respectively.
In the standard model, these two decay amplitudes contain $V_{ub}$ and loop-induced penguin operators~\cite{Buras:1998raa},
which could provide  weak and strong phases, respectively, resulting in non-vanishing direct CP-violating rate asymmetries (${\cal A}_{CP}$s).
Previous theoretical studies of ${\cal A}_{CP}$s in Eq.~(\ref{ACPexpt}) have been performed in various QCD models,
such as the perturbative  QCD (PQCD) method~\cite{pQCD}, 
generalized factorization approach~\cite{HG} and light-front quark model (LFQM)~\cite{WKL}.
These calculations in the literature are important to check   
if the results in the standard model are consistent with the experimental measurements.
In this work, we would like also to explore the CP-violating asymmetries with the MIT bag model.

It is known that to calculate the decay processes of the baryons, we need to know the details of the baryon wave functions. 
In the MIT bag model,  the quarks in the baryon are  confined in a  static bag. 
The model enjoys various successes in its simple structure  to explain the mass spectra and  magnetic dipole moments of the baryons~\cite{MIT_bag_1,MIT_bag_2,Bag48GeV,MIT_bag_3}.  However, the construction of the baryon wave functions in 
the MIT bag model is localized in a fixed space. 
As a result, the center motions of the baryons are not moved. In other words, the baryon wave functions are not in the momentum eigenstates. 
Such a defect makes the model questionable when one  discusses  the decay processes, involving essentially the momentum eigenstates. 
Nonetheless, many calculations in the bag model have  been done by ignoring this  problem~\cite{Boosting_the_bag,Lcwithbag,Isgur3,Cheng1,Cheng2,vectormeson,semi}. 
In this work,  we will modify the MIT bag model to construct the momentum eigenstates for the baryons to
study the decays without introducing extra parameters.

On the other hand, it is known that one can examine the heavy quark symmetry in
 the charmful decays of  $\Lambda_b^0 \to \Lambda_c^+M$.
 Particularly,  the baryonic matrix element is related to the Isgur Wise function of $\xi(\omega)$~\cite{Korner:1994nh}, given by
\begin{equation}
\langle \Lambda_c^+ | \overline{c} \gamma^\mu (1 - \gamma_5) b| \Lambda_b^0\rangle = 
	\overline{ u}_{\Lambda_c^+}  \xi (1 - \gamma_5) u_{\Lambda_b^0}\,,
\end{equation} 
where $\omega= v_1\cdot v_2$ with
$v_{1,2}$ the velocities of the two baryons and $u_{\bf B}$ is the 4 components Dirac spinor for the corresponding baryon of ${\bf B}$. 
In the heavy quark limit, in which  the relative velocity  between the two baryons vanishes, we have that $\xi(\omega) = 1 $ with $\omega= v_1\cdot v_2$. 
We will check if the heavy quark symmetry is valid in our modified MIT bag model

This paper is organized as follows. 
In Sec.~\MakeUppercase{\romannumeral 2}, we introduce the decay amplitudes and parameters.
In Sec.~\MakeUppercase{\romannumeral 3},
we constitute the baryon wave functions in the modified MIT bag model, where we sum over the localized baryon wave functions with different centers. In Sec.~\MakeUppercase{\romannumeral 4}, we compute the form factors accordingly. In Sec.~\MakeUppercase{\romannumeral 5}, we present our numerical results and compare them with the experimental data as well as the theoretical evaluations in the  literature. We conclude our study in Sec.~\MakeUppercase{\romannumeral 6}.

\section{Decay amplitudes and parameters}
We start with the  two-body decays of ${\bf B}_i \to {\bf B}_q M$, where ${\bf B}_{i(q)}$ is the initial (final) baryon with spin $1/2$
and $M$ represents the pseudoscalar meson.
In this study, we concentrate on the corresponding decays with ${\bf B}_i=\Lambda_b^0$, ${\bf B}_q=(\Lambda_c^+,p)$, and $M=(\pi^-,K^-)$.
The spin-dependent amplitude for ${\bf B}_i \to {\bf B}_q M$  can be written as 
\begin{equation}\label{general}
{\cal A} ({\bf B}_i \to {\bf B}_q M)= \overline{ u}_{{\bf B}_q} (A- B\gamma_5 ) u_{{\bf B}_i}\,,
\end{equation}
where $A$ and $B$ are the $s$-wave and $p$-wave amplitudes, corresponding to the parity violating and conserving ones,
and $u_{{\bf B}_{i,q}}$ are the baryon Dirac spinors, respectively.
In general, $A$ and $B$  are not relatively real, resulting in CP violating effects.
The decay branching ratio and  forward-backward asymmetry for the initial baryon in the rest frame are given as~\cite{ABanalyze} 
\begin{eqnarray}
&&\Gamma({\bf B}_i \to {\bf B}_q M)=\frac{|\vec{p}_q|}{8 \pi}\left(\frac{\left(M_{B_{i}}+M_{B_{q}}\right)^{2}-m_{M}^{2}}{M_{B_{i}}^{2}}|A|^{2}+\frac{\left(M_{B_{i}}-M_{B_{q}}\right)^{2}-m_{M}^{2}}{M_{B_{i}}^{2}}|B|^{2}\right)\,,\nonumber\\
&&\alpha_P(B_i \to B_q M)=\frac{2 \kappa \text{Re}\left(A^{*} B\right)}{|A|^{2}+\kappa^{2}|B|^{2}}, \quad \kappa=\frac{|\vec{p}_{q}|}{E_{B_{q}}+m_{B_{q}}}\,,
\end{eqnarray}
where $m_M$ is the meson mass and $\vec{p}_q$  represents the 3 momentum of the daughter baryon ${\bf B}_q$,
while  $\alpha_P$ describes the decay asymmetry between the helicity states  of  ${\bf B}_q$, defined by
\begin{equation}
\alpha_P ({\bf B}_i \to {\bf B}_q M)= \frac{ \Gamma(\hat{p}_q \cdot \hat{s}_q =1)-\Gamma(\hat{p}_q \cdot \hat{s}_q =-1)  }{\Gamma(\hat{p}_q \cdot \hat{s}_q =1)+\Gamma(\hat{p}_q \cdot \hat{s}_q =-1)}
\end{equation}
with $\hat{p}_q (\hat{s}_q)$ the unit vector in the $\vec{p}_q(\vec{s}_q)$  direction of  ${\bf B}_q$,  provided with  the initial baryon ${\bf B}_i$ unpolarized. 
We can also  define the decay asymmetries
 for the antiparticles, given by
\begin{eqnarray}
\overline{ \alpha }_P ({\bf B}_i \to {\bf B}_q M) &=&  \alpha_P (\overline{ \bf B}_i\to \overline{\bf B}_q  \overline{M} )\,.
\end{eqnarray}

The nonzero mean value between  $\alpha_P$ and $\overline{\alpha}_P$ is a consequence of CP violation,  
which is defined by~\cite{Brown:1983wd,Donoghue:1986hh}
\begin{equation}
\overline{ {\cal A} }_{CP}  ( {\bf B}_i \to {\bf B}_q M)  =\frac{\alpha_P  ( {\bf B}_i \to {\bf B}_q M)  + \overline{ \alpha }_P ( {\bf B}_i \to {\bf B}_q M)  }{\alpha_P  ( {\bf B}_i \to {\bf B}_q M)  - \overline{ \alpha }_P ( {\bf B}_i \to {\bf B}_q M)  }\,.
\end{equation}
On the other hand,  the direct
CP-violating rate asymmetries of the decays are given by 
\begin{equation}\label{directCP}
{\cal A}_{CP} ( {\bf B}_i \to {\bf B}_q M)  = \frac{\Gamma( {\bf B}_i \to {\bf B}_q M)   -\Gamma ( \overline{\bf B}_i \to \overline{\bf B}_q \overline{M})  }{\Gamma( {\bf B}_i \to {\bf B}_q M)   +\Gamma ( \overline{\bf B}_i \to \overline{\bf B}_q\overline{M})}\,.
\end{equation}

In the present work, to relate $\alpha_{P}$, ${\overline{\cal A}}_{CP}$ and ${\cal A}_{CP}$ directly to the weak interactions, 
we  ignored the  final states interactions in our calculations.
 As a result,  the complex phases of $A$ and $B$ solely come from the Wilson coefficients and the CKM elements.
We note that  possible sizable CP-violating effects in $\Lambda_b \to p(\pi^-, K^-)$
could be induced due to the weak phase from the CKM matrix element of $V_{ub}$ .

\section{Baryon wave functions}
In the  MIT bag model, the quarks in the baryon are  constrained in a certain bag with radius $R$. 
Inside the bag, each quark obeys the free Dirac equation, given by
\begin{equation}\label{free_dirac}
i\partial \!\!\!/ \psi - m_q \psi =0\,,
\end{equation}
due to the mean field approximation of the gluon field, where $m_q$ is the current quark mass. 
The boundary condition on the surface of the bag is given as 
\begin{equation}\label{boundary_condition}
in_i\partial_i \psi = \psi\,,
\end{equation}
where $n_i$ is the unit vector toward the surface.

In this work, we only consider the ground states of the baryons with their angular momenta to be $J=1/2$, where the spatial parts of 
the quark wave functions 
satisfy  the spherical symmetry.
Consequently, by inserting the boundary condition of Eq.~\eqref{boundary_condition} into Eq.~\eqref{free_dirac}, we get
\begin{equation}\label{momentum_condition}
\tan (p_qR) = \frac{p_qR}{1-m_qR-E_qR}\,,
\end{equation}
where $p_q$ is the magnitude of the 3-momentum for the quark and $E_q=\sqrt{p_q^2+m_q^2}$. The lowest momentum 
given in Eq.~\eqref{momentum_condition}  corresponds to the ground state. The quark wave function centered in $\vec{x}=0$  is written as 
\begin{equation}\label{quark_wave_function}
\psi(x) =\phi(\vec{x}) e^{-iE_qt}= \frac{N}{\sqrt{4\pi}} \left(
\begin{array}{c}
\omega_+ j_0(p_qr) \chi\\
i\omega_- j_1(p_qr) \hat{r} \cdot \vec{\sigma} \chi\\
\end{array}
\right)e^{-iE_qt}\,,
\end{equation}
where $\omega_\pm \equiv \sqrt{1\pm m_q/E_q}$, $\chi$ is the two-component spinor of the quark, describing the orientation of the angular momentum, 
 $j_i$ represents the $i$-th  spherical Bessel function of the first kind, and $N$ stands for the normalized factor, given by 
\begin{equation}
N=\sqrt{    \frac{ E_q(E_q-m_q)  }{ R^3j_0^2(p_qR )\left[2E_q\left(E_q-\frac{1}{R}\right) +\frac{m_q}{R}     \right]            }                       }\,.
\end{equation}
Since the baryon is  made of the product of the three quarks, we write its wave function in term of the quark wave functions $\psi_{q_i}$, given by
\begin{equation}\label{origin_wave}
\Psi(\vec{x}_{q_1},\vec{x}_{q_2},\vec{x}_{q_3},t) = \phi_{q_1}(\vec{x}_{q_1}) \phi_{q_2}(\vec{x}_{q_2}) \phi_{q_3}(\vec{x}_{q_3})e^{-i(E_{q_1}+E_{q_2}+E_{q_3})t}\,,
\end{equation}
where  $q_i$ ($i = 1, 2,3$)  denote the ith quarks and
$E_{q_i}$  are the corresponding energies.
Here, the spin, flavor and color configurations have not been written down explicitly.

However, the above construction of the baryon wave function is valid only around $\vec{x}=0$, which is mainly used to discuss 
the mass spectrum and magnetic dipole moment of the baryon. To calculate the dynamical processes, we need the  baryon state  to be in the particular 4-momentum eigenstate of the spacetime translation.

On the other hand, in  Eq.~\eqref{origin_wave} one assumes that the baryons are at rest, and the centers of the initial and final baryons locate in the same spot 
at a particular time, $t=t'$. However, if the baryons are not at relatively rest, the centers will be not coincident at $t\neq t'$ and the results of the dynamical factors will no longer be the same. Clearly, the baryon wave function in Eq.~\eqref{origin_wave} is not the eigenstate of the spacetime translation.

To construct the baryon wave function to be invariant under the space translation, we have to modify the MIT bag model. 
To do that,  we first integrate Eq.~\eqref{origin_wave} with different center locations, leading to 
\begin{equation}\label{wave_function_inrest}
\Psi(x_{q_1},x_{q_2},x_{q_3}) ={\cal N}\int 
d^3 \vec{x}
\prod_{i=1,2,3}
\phi_{q_i}(\vec{x}_{q_i} - \vec{x}) e^{-iE_{q_i}t_{q_i}} \,,
\end{equation}
where ${\cal N}$ is the normalization factor.
We then consider the space translation
\begin{equation}\label{translate_wave}
\Psi(x_{q_1}+d,x_{q_2}+d,x_{q_3}+d) ={\cal N}\int 
d^3 \vec{x}
\prod_{i=1,2,3}
\phi_{q_i}(\vec{x}_{q_i} +\vec{d}- \vec{x}) e^{-iE_{q_i}t_{q_i}}\,,
\end{equation}
where $d$ has only the spatial component.
By changing the integration variable $\vec{x}$ to $\vec{x}-\vec{d}$, we see that the wave function 
is indeed invariant under the space translation.

The energy of the baryon at rest essentially corresponds to the mass of the baryon, which can be read off from Eq.~\eqref{wave_function_inrest} to be $M=E_{q_1}+E_{q_2}+E_{q_3}$.
We note that the bag energy has not been included in our study.  In principle, one can calculate the energy contribution from 
the complicated gluon and gluon-quark interactions. 
However, we will not include such effects. Instead, we will simply use $M=E_{q_1}+E_{q_2}+E_{q_3}$ in the integral associated with the energy momentum conservation as an approximation.

The wave function for the baryon at rest is  given in Eq.~\eqref{wave_function_inrest}. To obtain the function in a certain 4-momentum, we boost it in the $z$-direction, given by 
\begin{eqnarray}
\label{WF}
\Psi_v(x_{q_1},x_{q_2},x_{q_3})&=&S^{\bf B}_v\Psi(x^v_{q_1},x^v_{q_2},x^v_{q_3}) ={\cal N} \int d^3\vec{x} \prod_{i=1,2,3} S^q_v\phi_{q_i}(\vec{x}^v_{q_i} -\vec{x})e^{-i\gamma E_{q_i}( t_{q_i} -v z_{q_i} )}\,.
\end{eqnarray}
Here, $S^{{\bf B}(q)}_v$ is the  pure Lorentz boost matrix for the baryon (quark) 4-spinor in the z-direction with $S^q_v$ given by
\begin{equation}\label{lorenz_boost}
S^q_v = 
\left(
\begin{array}{cc}
a_+ I &a_- \sigma_z\\
a_- \sigma_z &a_+ I
\end{array}
\right)\,,
\end{equation}
where $I$ is the $2\times2$ unity matrix, $a_\pm = \sqrt{\frac{1}{2}(\gamma\pm 1)}$ and $\gamma = \sqrt{1/(1-v^2)}$. 
In addition, $v$ in the superscript of $S^q_v$ indicates the Lorentz transformation of the coordinate, given explicitly
as 
\begin{eqnarray}\label{coordinate_LT}
&&(x^v_{q_i})_0 = \gamma (x_{q_i})_0 - \gamma v(x_{q_i})_3\nonumber\,,\\
&&(x^v_{q_i})_1 =(x_{q_i})_1 \,,~~~~~~(x^v_{q_i})_2 = (x_{q_i})_2 \nonumber\,,\\
&&(x^v_{q_i})_3=\gamma (x_{q_i})_3 - \gamma v(x_{q_i})_0\,,
\end{eqnarray}
where $(x_{q_i})_0$ is the time component of the coordinate.

To obtain the proper normalization factor, we calculate the overlap between the two baryon wave functions with different speeds 
at time $t$, given by
\begin{eqnarray}\label{overlapintegralwithitself}
&&\int \Psi^{\dagger}_{v'} (\vec{x}_{q_1}\,,\vec{x}_{q_2}\,,\vec{x}_{q_3},t) \Psi_v (\vec{x}_{q_1}\,,\vec{x}_{q_2}\,,\vec{x}_{q_3},t)d^3\vec{x}_{q_1} d^3\vec{x}_{q_2} d^3\vec{x}_{q_3}\nonumber\\
&& = e^{-iM(\gamma-\gamma')t}{\cal N}^2  \int d^3\vec{x}d^3\vec{x}'\prod_{i=1,2,3} d^3\vec{x}_{q_i} \phi^\dagger(\vec{x}^{v'}_{q_i} -\vec{x}')S^q_{v'} S^q_v \phi( \vec{x}^v_{q_i} -\vec{x}) e^{iE_{q_i} (\gamma v - \gamma' v' )  z_{q_i}  } \,,
\end{eqnarray}
where $S^{q\dagger }_v = S^q_v$ for the pure Lorentz boost and $M=E_{q_1}+E_{q_2}+E_{q_3}$ for the mass of the baryon. To simplify the integral, we adopt the following variables:
\begin{eqnarray}\label{variable_change}
&&\vec{x}^{ \,r}_{q_i} = \vec{x}^v_{q_i} - \frac{1}{2}(\vec{x}+\vec{x}')\,,\nonumber \\
&&\vec{x}_\Delta = \vec{x} - \vec{x}'\,,\nonumber\\
&&\vec{x}_A = \frac{1}{2}(\vec{x}+\vec{x}')\,.
\end{eqnarray}
Now, the overlap integral is read as 
\begin{eqnarray}
&&e^{-iM(\gamma-\gamma')t} \frac{{\cal N}^2}{\gamma^3}\int d^3\vec{x}_\Delta  d^3 \vec{x}_A\prod_{i=1,2,3}  d^3 \vec{x}^{\,r}_{q_i} \phi^\dagger(\vec{x}^{ \,r}_{q_i} +\frac{1}{2}\vec{x}_\Delta) S^{q\,2}_v \psi(\vec{x}^{ \,r}_{q_i}  -\frac{1}{2}\vec{x}_\Delta) e^{iE_{q_i}^B(v-v')z^r_{q_i}} e^{iE_{q_i}( v-v'   )z_A  } \nonumber\\
&&={\cal N}^2\gamma(2\pi)^3\delta^3\left(\vec{p}-\vec{p}'\right) \int d^3\vec{x}_\Delta\prod_{i=1,2,3} d^3 \vec{x}_{q_i}^{\,r} \phi^\dagger\left(\vec{x}^{\,r}_{q_i}+ \frac{1}{2} \vec{x}_\Delta \right)\phi\left(\vec{x}^{\,r}_{q_i} - \frac{1}{2} \vec{x}_\Delta \right)\,,
\end{eqnarray}
where $1/\gamma^3$ comes from the Jacobian in Eq.~\eqref{variable_change}, and $\vec{p}$ and $\vec{p}'$ are the 3-momenta of the baryons. 
Here, we have used the fact that the integral does not vanish if and only if $v=v'$ to reduce the complexity  in $\phi$.

By normalizing the baryon wave function as $\langle p' | p \rangle = \gamma (2\pi)^3 \delta^3( \vec{p} -\vec{p}')$, we find that
\begin{equation}\label{normalizaiton}
\frac{1}{{\cal N}^2 } = \int d^3\vec{x}_\Delta\prod_{i=1,2,3} d^3 \vec{x}_{q_i}^{\,r} \phi^\dagger\left(\vec{x}^{\,r}_{q_i} + \frac{1}{2} \vec{x}_\Delta \right)\phi\left(\vec{x}^{\,r}_{q_i}- \frac{1}{2} \vec{x}_\Delta \right)\,,
\end{equation}
 which is clearly independent of the velocity by the construction of the baryon wave functions.

It is worthwhile to take a look at the physical interpretation of the baryon wave function in Eq.~\eqref{wave_function_inrest}. 
In contrast to the original wave function in Eq.~\eqref{origin_wave} with the quarks confined in a static bag located at $\vec{x}=0$, 
those quarks for Eq.~\eqref{wave_function_inrest} distribute all over the space to fulfill the translation-invariant requirement. Note that the distance between two arbitrary quarks is limited within $2R$ in the bag model. For the case in Eq.~\eqref{wave_function_inrest}, it is due to that if $|\vec{x}_{q_1} - \vec{x}_{q_2}| >2R$, then either $|\vec{x}_{q_1}- \vec{x} |$ or $|\vec{x}_{q_2} - \vec{x} |$  will exceed $R$, resulting in the  vanishing integral. So the quarks are entangling to each other in the spatial part of the wave function, which is not the case in Eq.~\eqref{origin_wave}.  

The average distance between the quarks is given by 
\begin{eqnarray}
\sqrt{ \langle \left(\vec{x}_{q_1} -\vec{x}_{q_2}\right)^2 \rangle} = \sqrt{\langle \vec{x}_{q_1}^2 \rangle
+ \langle \vec{x}_{q_2}^2 \rangle - 2 \langle \vec{x}_{q_1} \cdot \vec{x} _{q_2} \rangle }\,.
\end{eqnarray}
In the original MIT bag model, as the quark positions are independent to each other, we have that $\langle \vec{x}_{q_1} \cdot \vec{x} _{q_2} \rangle  = \langle \vec{x}_{q_1} \rangle \cdot \langle \vec{x} _{q_2} \rangle  = 0$. However, they do not vanish in our modified MIT bag model. For the proton with 3 massless light quarks, the average distance in our modified model is roughly $20\%$ shorter than the original one with the same bag radius.

\section{Baryon form factors}
To calculate the $\Lambda_b^0$ decays, we use the factorization approach. In this approach, the amplitudes for $\Lambda_b^0 \to \Lambda^+_c M$ 
can be written as 
\begin{eqnarray}
\label{Ac}
{\cal A}_{(\Lambda_b^0 \to \Lambda_c^+ M)} &=& \frac{G_F}{\sqrt{2}} a_1 V_{cb}^* V_{uq} \langle M | \overline{u} \gamma^\mu(1-\gamma_5) q|0\rangle \langle \Lambda_c^+ | \overline{c}\gamma_\mu (1-\gamma_5) b | \Lambda_b^0\rangle  \nonumber\\
&=& i \frac{G_F}{\sqrt{2}} a_1 V_{cb}^* V_{uq} f_M q^\mu \langle \Lambda_c^+ | \overline{c} \gamma_\mu ( 1 - \gamma_5) b | \Lambda_b^0\rangle \nonumber\\
&=& i \frac{G_F}{\sqrt{2}} a_1 V_{cb}^* V_{uq} f_M \left[  
(m_b - m_c) \langle \Lambda_c^+ | \overline{c}b |\Lambda_b^0\rangle + (m_b + m_c) \langle \Lambda_c^+ | \overline{c }\gamma_5 b |\Lambda_b^0\rangle
\right]\,,
\end{eqnarray}
where  $G_F$ is the Fermi constant, $a_1= c_1 + c_2/3=1.02$~\cite{Buras:1998raa} with $c_{1,2}$ the Wilson coefficients, 
$V_{ij}$ represent the CKM elements with $q=d(s)$ corresponding to $M=\pi^-(K^-)$, 
$f_M$  is the meson decay constant, and the quarks operators are evaluated at $x=0$. 
For the decay of $\Lambda_b^0 \to p M$\,, the amplitude is given by~\cite{HG}
\begin{equation}
\label{Au}
{\cal A} (\Lambda_b^0 \to pM) =i \frac{G_F}{\sqrt{2}}m_b f_M\left[   
\alpha_M \langle p | \overline{u} b | \Lambda_b^0 \rangle + \beta_M \langle p | \overline{u} \gamma_5 b|\Lambda_b^0 \rangle 
\right]\,,
\end{equation}
where  $\alpha_{M}$ ($\beta_M$) and $\alpha_{V}$ in Eq.~(\ref{eq1}) are defined by 
\begin{eqnarray}\label{alphaBeta}
\alpha_{M}&=& V_{ub}V_{uq}^*a_1-V_{tb}V_{tq}^*(a_4+ r_M a_6)\;,\nonumber\\
\beta_{M}&=& V_{ub}V_{uq}^*a_1-V_{tb}V_{tq}^*(a_4- r_M a_6),
\end{eqnarray}
with $r_M\equiv {2 m_M^2}/[m_b (m_q+m_u)]$
and  $a_i\equiv c^{eff}_i+c^{eff}_{i\pm1}/N_c^{(eff)}$ for $i=$odd (even),
 composed of the effective Wilson coefficients $c_i^{eff}$ defined in Ref.~\cite{ali}.

Now, we are left with the matrix elements of the scalar and pseudoscalar operators in Eqs.~(\ref{Ac}) and (\ref{Au}), which can be parametrized as 
\begin{eqnarray}\label{eq0}
\langle {\bf B}_q | \overline{ q } b (0)| \Lambda_b^0 \rangle &=& f_s^{{\bf B}_q} \overline{u }_{{\bf B}_q} u_{\Lambda_b^0}
\nonumber\\
\langle {\bf B}_q | \overline{ q }\gamma_5  b (0)| \Lambda_b^0 \rangle &=& f_p^{{\bf B}_q} \overline{u }_{{\bf B}_q} \gamma_5 u_{\Lambda_b^0}\,,
\end{eqnarray}
where ${\bf B}_q$ represents $\Lambda_c^+(p)$ with $\overline{q}$  being $\overline{c}(\overline{u})$.

We evaluate the form factors of $f_s^{{\bf B}_q}$ and $f_p^{{\bf B}_q}$ in the Briet frame, in which the initial and final baryons have opposite velocities, 
i.e. $\vec{v}_{1,2}=-\vec{v},\,\vec{v}$.
In the derivations of the matrix elements, one actually deals  with the quark operators in the $x$-dependence, given by 
\begin{equation}\label{eq1}
\int \langle {\bf B}_q | \overline{q}(\gamma_5) b(x) e^{ip_Mx} | \Lambda_b^0\rangle d^4x =\langle {\bf B}_q | \overline{q} (\gamma_5)b(0) | \Lambda_b^0\rangle (2\pi)^4 \delta( p_i-p_q-p_M) \,,
\end{equation}
for the scalar (pseudoscalar) matrix element with $p_M$  being the 4-momentum of the $M$ meson. 
Here, we have used  that the initial and final baryons are in the momentum eigenstates to reduce the integral with
the  Dirac $\delta$ function of  $(2\pi)^4 \delta( p_i-p_q-p_M)$.
Clearly, one can evaluate the form factors with either the quark operators located at  $x=0$ in  Eq.~\eqref{eq0} or  the $x$-dependent ones
 in the left hand side of Eq.~\eqref{eq1}.
We start  with the baryon wave functions in Eq.~(\ref{WF}),
 given by
\begin{eqnarray}
&&\int \langle {\bf B}_q | \overline{ q } b(x_{q_3}) e^{ip_Mx_{q_3}} |\Lambda_b \rangle d^4 x_{q_3}\nonumber\\
&=& {\cal N}_{\Lambda_b} {\cal N}_{B_q}\int d^3 \vec{x} d^3 \vec{x}' d^4x_{q_3} \overline{ \phi }_{q} (\vec{x}^v_{q_3} -\vec{x}') S_{-v}^{q\,2}\phi_b(\vec{x}^{-v}_{q_3} -\vec{x} ) e^{i\left[\gamma(E_q-E_b) +p_0\right]t}\nonumber\\
&&e^{-i[\gamma v(E_q+E_b)+p_3]z_3 }\prod_{j=1,2} \phi_{q_j}^\dagger(\vec{x}^v_{q_j} - \vec{x}')  \phi_{q_j}(\vec{x}^{-v}_{q_j} - \vec{x}) e^{-2i\gamma v E_{q_j} v z_{q_j} }\,,
\end{eqnarray}
with $S^{q\dagger}_v \gamma_0 = \gamma_0 S^{q}_{-v} $  for the Lorentz boost and $(q_1,q_2) = (u,d)$. 
Similar to the case in Eq.~\eqref{overlapintegralwithitself}, we adopt the variable transformations with some modifications, given by
\begin{eqnarray}
&&(\vec{x}_{q_i}^{\,r})_j=(\vec{x}_{q_i})_j -\frac{1}{2}(\vec{x}' +\vec{x}) _j \,,\nonumber\\
&&(\vec{x}^{\,r}_{q_i})_{3} = \gamma (\vec{x}_{q_i} )_3 -\frac{1}{2}(\vec{x}' +\vec{x}) _3\,,\nonumber\\
&&\vec{x}_A = \frac{1}{2}(\vec{x}+\vec{x}')\,,\nonumber\\
&&(\vec{x}_\Delta)_j = (\vec{x} - \vec{x}')_j\,,\nonumber\\
&&(\vec{x}_\Delta)_3= (\vec{x} - \vec{x}')_3 -2 \gamma v t\,,
\end{eqnarray}
where $i=1,2,3$ and $j=1,2$.
Subsequently, we have that
\begin{eqnarray}\label{master}
&&{\cal Z}\int d^3\vec{x}_\Delta d^3\vec{x}^{\,r}_{q_3} \overline{ \phi }_q\left(\vec{x}^{\,r}_{q_3} + \frac{1}{2}\vec{x}_\Delta \right)S^{q\,2}_{-v} \phi_b\left(\vec{x}^{\,r}_{q_3} - \frac{1}{2}\vec{x}_\Delta \right) e^{iv(M_{B_f} + M_{\Lambda_b}-E_q-E_b)z^{\,r}_{q_3}     }\prod_{j=1,2} D_{q_j}(\vec{x}_\Delta)\,,\nonumber\\
&&{\cal Z} \equiv (2\pi )^4 \delta^4(p_i - p_f -q_M)    \frac{{\cal N}_{\Lambda_b} {\cal N}_{B_f} }{\gamma^2}\,,\nonumber\\
&&D_{q_j}(\vec{x}_{\Delta}) \equiv \int d^3 \vec{x} \phi_{q_j}^\dagger \left(\vec{x} +\frac{1}{2}\vec{x}_{\Delta}\right) \phi_{q_j} \left(\vec{x} -\frac{1}{2}\vec{x}_{\Delta}\right)
e^{-2iE_{q_j}  v z_{q_j} }\,.
\end{eqnarray}

Alternatively, one can evaluate the integral with  $b$ and $q$ quarks  located at $x=0$  in Eq.~\eqref{eq0}\,, given as
\begin{equation}
{\cal Z} \gamma^2\int d^3 \vec{x} d^3 \vec{x}' \overline{ \phi }_q(-\vec{x}') S_{-v}^{q\,2} \phi_b (-\vec{x})\prod_{j=1,2} \phi_{q_i}^\dagger(\vec{x}^v_{q_i} - \vec{x}')  \phi_{q_i}(\vec{x}^{-v}_{q_i} - \vec{x}) e^{-2i\gamma E_{q_j} v z_{q_j} }\,.
\end{equation}
After changing the integral variables by
\begin{eqnarray}
&&(\vec{x}^r_{q_j})_{k} = (\vec{x}_{q_j})_{k} - \frac{1}{2} (\vec{x} + \vec{x}' )_k\,,\nonumber\\
&&(\vec{x}^{\,r}_{q_j})_{3} = \gamma (\vec{x}_{q_j} )_3 -\frac{1}{2}(\vec{x}' -\vec{x}) _3\,,\nonumber\\
&&\vec{x}_B = -\frac{1}{2}\left( \vec{x} +\vec{x}' \right)\,, \nonumber\\
&& \vec{x}_{\Delta} = \vec{x} -\vec{x}'\,,
\end{eqnarray}
where $k, j = 1,2$\,,
 one obtains the identical equation as the one in  Eq.~\eqref{master}
 with $M= E_{q_1} +E_{q_2} + E_{q_3}$.

Similarly, the pseudoscalar part can be given as 
\begin{eqnarray}
&&\int \langle {\bf B}_q | (\overline{q}\gamma_5 b)(x) e^{ip_Mx} | \Lambda_b^0\rangle d^4x= \\
&&{\cal Z}\int d^3\vec{x}_\Delta d^3\vec{x}^{\,r}_{q_3} \overline{ \phi }_q\left(\vec{x}^{\,r}_{q_3} + \frac{1}{2}\vec{x}_\Delta \right)\gamma_5S^{q\,2}_{-v} \phi_b\left(\vec{x}^{\,r}_{q_3} - \frac{1}{2}\vec{x}_\Delta \right) e^{iv(M_{{\bf B}_q} + M_{\Lambda_b}-E_q-E_b)z^{\,r}_{q_3}     }\prod_{j=1,2} D_{q_j}(\vec{x}_\Delta)\,.\nonumber
\end{eqnarray}
With the normalization in Eq.~\eqref{normalizaiton}, we derive that 
\begin{eqnarray}\label{fsfp}
f_s&=&\frac{\eta}{\gamma(2\pi)^4  \delta^4(p_i-p_f-p)}\int \langle {\bf B}_q | (\overline{q} b)(x) e^{ip_Mx} | \Lambda_b^0\rangle d^4x\,,\nonumber\\
f_p&=&\frac{\eta}{\gamma(2\pi)^4  \delta^4(p_i-p_f-p)}\int \langle {\bf B}_q | (\overline{q}\gamma_5 b)(x) e^{ip_Mx} | \Lambda_b^0\rangle d^4x\,,
\end{eqnarray}
where $\eta$ is the overlap factor of the spin-flavor configuration. For $\Lambda_b \to \Lambda_c^+ (p)$, we have $\eta = 1~ ( \sqrt{3/2})$~\cite{Lcwithbag}.

\section{Numerical results and discussions}
We use the bag radius of the $\Lambda_c^+$ baryon as $R(\Lambda_c^+)=4.8$~GeV$^{-1}$ from Ref.~\cite{Bag48GeV}. 
In the limit of the heavy quark symmetry,  the baryon wave functions for $\Lambda_b^0$ and $\Lambda_c^+$ 
can be taken to be the same.
As a result,  the bag radius for the $\Lambda_b^0$ should be also around 4.8~GeV. 
For the proton, it is usually chosen to be $5$~GeV$^{-1}$. However, to simplify our numerical calculations,
 we choose  the  same bag radiuses for   $\Lambda_b^0$, $\Lambda_c^+$ and $p$, $i.e.$ 
  $R=R(\Lambda_b^0)=R(\Lambda_c^+)=R(p)=4.8\pm0.2$~GeV$^{-1}$,
where the uncertainty is to account for the dependence of the bag radius.
In general, an increase of  the bag radius would reduce the  form factors.
In addition, we take that $m_u=m_d=5$~MeV.
Note that the variation of the light quark masses from $0$ to $10$~MeV makes no much difference for the numerical values of the form factors.
To determine the heavy quark masses, we assume  that the baryon mass differences are related to   the corresponding quark energies, i.e.
\begin{eqnarray}
E_b = M_{\Lambda_b^0} - M_p + E_u\,,\nonumber\\
E_c = M_{\Lambda_c^+} - M_p + E_u\,,
\end{eqnarray}
where $E_b$, $E_c$ and $E_u$ are the energies of $b$, $c$ and $u$ quarks in the bag, and $M_{\Lambda_b^0}$, $M_{\Lambda_c^+}$ and $M_p$ are the corresponding baryons masses, respectively. With $R=4.8$~GeV$^{-1}$, we find $(m_b\,,m_c ) = (5.1\,,1.8)$~GeV, which satisfy the relation of the heavy quark symmetry, given by $m_b - m_c = M_{\Lambda_b^0} - M_{\Lambda_c^+}$.

The  form factors associated with the vector and axial-vector currents are defined by
\begin{eqnarray}
\label{eqVA}
\langle {\bf B}_q | \overline{ q } \gamma_\mu b |\Lambda_b^0 \rangle& =& \overline{u }_f 
\left(  f_1 \gamma_\mu -f_2 i \sigma_{\mu \nu} (p_M)^\nu +f_3 (p_M)_\mu  
\right) u_{\Lambda_b^0}\,,\nonumber\\
\langle {\bf B}_q | \overline{ q } \gamma_\mu \gamma_5 b |\Lambda_b^0 \rangle &=& \overline{u }_f 
\left(  g_1 \gamma_\mu -g_2 i \sigma_{\mu \nu} (p_M)^\nu +g_3 (p_M)_\mu  
\right) \gamma_5 u_{\Lambda_b^0}\,.
\end{eqnarray}
By using the equations of motion and comparing Eqs.~(\ref{eq0}) and (\ref{eqVA}), we obtain the form factors associated with the vector and axial currents from the scalar and pseudoscalar ones, given by
\begin{eqnarray}\label{eqmotion}
f_1 = (m_b - m_q )/(M_{\Lambda_b^0} - M_{{\bf B}_q} )  f_s\,,\nonumber\\
g_1 = (m_b + m_q )/(M_{\Lambda_b^0} + M_{{\bf B}_q} )  f_p\,,
\end{eqnarray}
respectively.
Here, we have   neglected the contributions from $f_3$ and $g_3$\,, which are suppressed based on the heavy baryon mass of $\Lambda_b$. 
Due to the relation of  $m_b - m_c = M_{\Lambda_b^0} - M_{\Lambda_c^+}$, we find that $f_s^{\Lambda_c^+}=f_1^{\Lambda_c^+}$, 
whereas $f_p$ is larger than $g_1$ in all cases.

Our numerical results of the form factors with different values of $p_M^2$ are listed in Table~\ref{Table1},
where we only consider the uncertainties  from the bag radius.
\begin{table}
	{\footnotesize
	\caption{Form factors for $\Lambda_b^0 \to \Lambda_c^+/ p$.
	}
		\begin{tabular}[t]{|c|cccc|cccc|}
\hline
$p_M^2 $&$f_s^{ \Lambda_c^+}$&$f_p^{ \Lambda_c^+}$&$f_1^{ \Lambda_c^+}$&$g_1^{ \Lambda_c^+}$&$f_s^{ p}$&$f_p^{ p}$&$f_1^{p}$&$g_1^{p}$\\
\hline
\hline
$(M_{\Lambda_b^0 }-M_{\Lambda_c^+ })^2 $ &$1.02\pm 0.00$&$1.20\pm 0.00$  &  $1.02\pm 0.00$&$1.04\pm 0.01$  &$1.32\pm 0.00$ &$2.10\pm 0.04$ &$1.44\pm 0.00$  & $1.64\pm 0.03$  \\
$M_\pi^2$ & $0.50\pm 0.00$& $0.58\pm 0.01$ &  $0.50 \pm 0.00$ &$0.50\pm 0.00$  &   $0.14\pm 0.01$  & $0.18\pm 0.01$  & $0.16\pm0.01$ & $0.14\pm0.01$    \\
$ M_K^2$&$0.51\pm 0.01$ &$0.59\pm 0.01$    &   $0.51\pm 0.01$&$0.51\pm 0.00$   &  $0.14\pm 0.01$  & $0.18\pm 0.01$  & $ 0.16\pm0.01$  & $0.14\pm0.00$   \\
\hline
		\end{tabular}
	\label{Table1}
}
\end{table}
For $\Lambda_b^0 \to \Lambda_c^+$,  with the heavy quark symmetry limit we have that
\begin{eqnarray}\label{heavy_quark}
&&f_1^{ \Lambda_c^+}(p_M^2)= g_1^{ \Lambda_c^+}(p_M^2)\,,\nonumber\\
&&f_1^{ \Lambda_c^+}\left(p_M^2 = (M_{\Lambda_b^0 }-M_{\Lambda_c^+ })^2\right) = g_1^{ \Lambda_c^+}\left(p_M^2 = (M_{\Lambda_b^0 }-M_{\Lambda_c^+ })^2\right) = 1\,,
\end{eqnarray}
which are  well consistent with our numerical values. 
The  results with the heavy quark symmetry indicate that our constructions for baryon wave functions are reasonable, while  the form factors
of $f_3^{\Lambda_c^+}$ and $g_3^{\Lambda_c^+}$ can be safely neglected.
It is interesting to note that
$f_1^{\Lambda_c^+}$ and $g_1^{\Lambda_c^+}$ correspond to the Isgur Wise function in the context of the heavy quark symmetry
The first derivatives of $f_1^{\Lambda_c^+}$ and $g_1^{\Lambda_c^+}$ are found to be
\begin{eqnarray}
\label{rhoVA}
\rho_V^2 = - \left.\frac{d}{d \omega} f_1(\omega) \right| _{\omega = 1} &=& 1.96\pm 0.04\,,\nonumber\\
\rho_A^2 = - \left.\frac{d}{d \omega} g_1(\omega) \right| _{\omega = 1} &=& 2.07\pm 0.04\,
\end{eqnarray}
where the slightly difference between $\rho_V^2$ and $\rho_A^2$ can be viewed as the $(1/m_b)$ correction.
Our results in  Eq.~(\ref{rhoVA}) are consistent with $\rho^2 = \rho_V ^2=\rho_A^2= 1.3 \-- 3.7$
in the literature based on the heavy quark symmetry~\cite{Isgur1,Isgur2,Isgur3,Isgur4}.     
On the other hand, the experimental measurement on  $\Lambda_b^0 \to \Lambda_c^+ \ell^-\bar{\nu}_\ell$ gives~\cite{IsgurExp}
\begin{equation}
\rho^2 = 2.04 \pm 0.46~ \text{(stat)}~ ^{+0.72}_{-1.00}~ \text{(syst)}\,,
\end{equation}
in which the central value is very close to our values in Eq.~(\ref{rhoVA}).

The  decay widths and asymmetries are shown in Table~\ref{decay_parameters_table}. 
As $f_1=g_1$ from the heavy quark symmetry   and $m_M^2\approx 0$  due to the  soft meson limit,
we expect that $A = \kappa B$, leading to   $\alpha_P(\Lambda_b^0\to \Lambda_c^+\pi^-,  \Lambda_c^+K^-)= 1$, as given in the table.
\begin{table}
	\caption{Decay widths and symmetries.
	}
		\begin{tabular}[t]{|c|ccccc|}
		\hline
channel &$ \Gamma (\text{s}^{-1})$ & ~$\alpha_P $(\%)& ~$\overline{\alpha}_P $(\%) &~$\overline{{\cal A}}_{CP}$(\%)&~${\cal A}_{CP}$(\%) \\
\hline
\hline
$\Lambda_b^0 \to \Lambda_c^+ \pi^-$ & $ ( 3.03 \pm 0.07 )\times 10^9$& $100.0\pm 0.0$ &$-100.0\pm 0.0$ &  0& 0\\
$\Lambda_b^0 \to \Lambda_c^+ K^-$ & $ (2.33\pm 0.06)\times 10^8$& $100.0\pm 0.0$ &$-100.0\pm 0.0$& 0& 0 \\
$\Lambda_b^0 \to p \pi^-$ & $ (3.41\pm 0.38)\times 10^6$& $85.6\pm 0.0$ & $-83.2\pm 0.1$  &$1.4\pm 0.0$& $-4.4\pm 0.1 $\\
$\Lambda_b^0 \to p K^-$ & $ (4.11\pm 0.49) \times 10^6 $& $-29.7\pm 0.3$  & $ 44.4\pm 0.4$ & $-19.6\pm 0.2 $&$6.7\pm 0.0$\\
\hline
		\end{tabular}
	\label{decay_parameters_table}
\end{table}
In Table~\ref{compared}, 
\begin{table}
	\caption{Decay branching ratios and direct CP asymmetries.}
	\begin{tabular}[t]{|c|cccccc|}
		\hline
		&  Our results& \cite{HG}&\cite{WKL}& \cite{pQCD} & \cite{Chua} & PDG~\cite{pdg}\\
		\hline
		\hline
		$10^3{\cal B} (\Lambda_b^0 \to \Lambda_c^+ \pi^-)$& $4.5\pm 0.2$ &-&-&-&$4.16^{+2.43}_{-1.73}$&$4.9\pm 0.4 $\\
		$10^4{\cal B} (\Lambda_b^0 \to \Lambda_c^+ K^-)$&$ 3.4\pm 0.1$&-&-&-&$ 3.1 ^{+ 1.8}_{-1.3}$ &$ 3.6 \pm  0.3$ \\
		$10^6{\cal B} (\Lambda_b^0 \to p \pi^-)$& $5.0\pm 0.5 $ &$4.2\pm 0.7$&$4.30$&$ 5.2^{+2.5}_{-1.9} $& -&$ 4.5 \pm 0.8 $\\
		$10^6{\cal B} (\Lambda_b^0 \to p K^-)$&$6.0 \pm 0.7$&$ 4.8 \pm 0.7 $&$2.17$&$2.0^{+1.0}_{-1.3}$&-&$5.4\pm 1.0$ \\
		$10^2{\cal A}_{CP} (\Lambda_b^0 \to p \pi^-)$& $-4.4\pm 0.1 $& $-3.9\pm 0.2$&$-3.37^{+0.29}_{-0.37}$&$-31^{+43}_{-1}$&-&$-2.5\pm 2.9$\\
		$10^2{\cal A}_{CP} (\Lambda_b^0 \to p K^-)$&$ 6.7\pm 0.0$& $ 5.8\pm 0.2    $&$10.1^{+1.3}_{-2.0}$&$-5^{+26}_{-5}$&-&$-2.5\pm 2.2$\\
		\hline
	\end{tabular}
	\label{compared}
\end{table}
we compared our results with those of Refs.~\cite{HG,WKL,pQCD,Chua}
in the literature as well as the experimental data~\cite{pdg}. 
In the literature,
 the form factors are evaluated by fitting  the experimental data in the generalized factorization approach~\cite{HG},
 considering  the LFQM  for the baryon wave functions~\cite{WKL,Chua}, 
  and 
 using the perturbative QCD method with the hybird scheme~\cite{pQCD}. 
 As shown in Table~\ref{compared},
 the decay branching ratios for $\Lambda_b^0\to \Lambda_c^+(\pi^-,K^-) $ 
 from the modified bag model are close to those in LFQM~\cite{Chua}  as well as the experimental data.
 We also find that our predicted  branching ratio for $\Lambda_b^0\to pK^- $ 
 is about 1.2 times larger than that for  $\Lambda_b^0\to p\pi^- $, which agrees with the data and that in the generalized factorization approach~\cite{HG},
 but  different from the results of pQCD~\cite{pQCD} and LFQM~\cite{WKL}.
 On the other hand, our results for the direct CP-violating rate asymmetries of  $\Lambda_b^0\to p(\pi^-,K^-) $ are sizable, which  are consistent with
  all other theoretical predictions, and the experimental data  except ${\cal A}_{CP} (\Lambda_b^0 \to p K^-)_{\rm PDG}$.
 As the experimental value of  ${\cal A}_{CP} (\Lambda_b^0 \to p K^-)_{\rm PDG}$ in Eq.~(\ref{ACPexpt})
 is consistent with zero with a negative central value, whereas our prediction of +6.7\% along with the others in Refs.~\cite{HG,WKL}
 is positive, it is very interesting to see
 if such CP  asymmetry can be measured precisely by the ongoing experiment at LHCb.
 In addition, we see that $\overline{{\cal A}}_{CP}(\Lambda_b^0 \to p K^-)$ is predicted to be $(-19.6 \pm 0.2) \%$, which is very large.
  
In Table~\ref{Table4}, we illustrate the ratio of $R={\cal B}(\Lambda_b^0\to \Lambda_c^+ K^-)/{\cal B}(\Lambda_b^0\to \Lambda_c^+ \pi^-)$
in various approaches.
\begin{table}
	\caption{Values  ($10^{-2}$) of $R={\cal B}(\Lambda_b^0\to \Lambda_c^+ K^-)/{\cal B}(\Lambda_b^0\to \Lambda_c^+ \pi^-)$ 
in various approaches.}
	\begin{tabular}[t]{|c|c|c|c|c|c|}
		\hline
	  Our result & U-spin  & Factorization  &   LFQM~\cite{Chua} & LHCb~\cite{piKratio} & PDG~\cite{pdg}\\
		\hline
		\hline
		$7.6\pm 0.1$ & 5.3 & 7.7 & $7.5$ & $7.21\pm0.22$ &  $7.35\pm 0.86 $\\
		\hline
	\end{tabular}
	\label{Table4}
\end{table}
In the table, 
 the result of the U-spin symmetry is based on the $SU(2)$ symmetry between $d$ and $s$  quarks, which 
 leads to  the naive relation for $R$, given by
\begin{equation}
R_{\text{U-spin}}=\left|\frac{V_{us}}{V_{ud}}\right|^2 \approx 5.3\%\,.
\end{equation}
In the factorization approach,  $R$ receives an extra factor due to the meson decay constants, read as
\begin{equation}\label{ratio}
R_{\text{Factorization}}=\left|\frac{V_{us}f_K}{V_{ud}f_\pi}\right|^2 \approx 7.7\%\,,
\end{equation}
which is consistent with our result and that in LFQM~\cite{Chua} as well as the data~\cite{piKratio,pdg}.
Clearly,  it shows the evidence that the  decays of $\Lambda_b^0\to \Lambda_c^+(\pi^-,K^-) $ are factorizable.

\section{Conclusions}
We have studied the decays of $\Lambda_b^0 \to \Lambda_c^+ (\pi^-,  K^-)$ and $\Lambda_b^0 \to p (\pi^-,K^-)$
in the modified MIT bag model.
We have provided a new way to construct the baryon  momentum eigenstates in the bag model without introducing new parameters. 
In particular,  we have summed over the localized baryon wave function in Eq.~\eqref{origin_wave} with different centers
to fulfill the requirement of  the invariant for the space  translation. 

For  $\Lambda_b^0 \to \Lambda_c^+(\pi^- , K^-)$, we have found that the decay branching ratios are
$(4.5\pm 0.2)\times 10^{-3}$ and $(3.4\pm 0.1)\times 10^{-4}$
 with the uncertainties only from the bag radius,
 which agree well with the experimental data of 
$(4.9\pm 0.4) \times 10^{-3}$ and $(3.6\pm 0.3)\times 10^{-4}$, respectively. We have also shown that our results of the first derivatives for the 
form factors $f_1(\omega)$ and $g_1(\omega)$ in Eq.~(\ref{rhoVA}) match with the data as well as those in the literature,
indicating the validation of the heavy quark symmetry in the decay processes.

For $\Lambda_b^0 \to p (\pi^-,K^-)$, our predicted decay branching ratios  of $(5.0\pm 0.5) \times 10^{-6}$ and $(6.0\pm 0.7) \times 10^{-6}$
are consistent with the current data of $(4.5\pm0.8)\times 10^{-6}$ and $(5.4\pm 1.0)\times 10^{-6}$~\cite{pdg}, respectively.
In addition, we have explored the CP-violating  asymmetries for the decays.
Particularly, we have obtained that ${\cal A}_{CP}(\Lambda_b^0 \to p \pi^-)$ and ${\cal A}_{CP}(\Lambda_b^0 \to p K^-)$ are $(-4.4\pm 0.1)\%$  
and $(6.7\pm 0.0)\%$,  in comparison with $(-2.5\pm2.9)\%$ and $(-2.5\pm2.2)\%$ from the Particle Data Group in 2020,  respectively.
It is also interesting to note that $\overline{{\cal A}}_{CP}(\Lambda_b^0 \to p K^-)$ is predicted to be $(-19.6\pm 0.2) \%$, which is very large.
It is clear that more precise future experimental measurements on these CP violating asymmetries are needed.

\section*{ACKNOWLEDGMENTS}
This work was supported in part by National Center for Theoretical Sciences and
MoST (MoST-107-2119-M-007-013-MY3).


\begin{thebibliography}{99}

\bibitem{Aaij:2014lpa} 
R.~Aaij {\it et al.} [LHCb Collaboration],
JHEP {\bf 1404}, 087 (2014). 

\bibitem{Aaltonen:2008hg} 
T.~Aaltonen {\it et al.} [CDF Collaboration],
Phys.\ Rev.\ Lett.\  {\bf 103}, 031801 (2009). 


\bibitem{Aaij:2015fea}
R.~Aaij {\it et al.} [LHCb Collaboration], Chin.\ Phys.\ C {\bf 40}, 011001 (2016).

\bibitem{Aaij:2016ymb}
R.~Aaij {\it et al.} [LHCb Collaboration],
Phys.\ Rev.\ Lett.\  {\bf 117}, 082003 (2016).


\bibitem{LHCbDcharmB}.
R.~Aaij {\it et al.} [LHCb Collaboration],
%
Phys.\ Rev.\ Lett.\  {\bf 119}, 112001 (2017).

\bibitem{CPnew}
R.~Aaij \textit{et al.} [LHCb Collaboration],
Phys. Lett. B \textbf{787}, 124 (2018).

\bibitem{pdg}
P.A. Zyla et al. (Particle Data Group), Prog. Theor. Exp. Phys. 2020, 083C01 (2020).

\bibitem{Buras:1998raa}
A.~J.~Buras,
[arXiv:hep-ph/9806471 [hep-ph]].


\bibitem{pQCD}
C.~D.~L$\ddot{\text{u}}$ , Y.~M.~Wang, H.~Zou, A.~Ali and G.~Kramer,
Phys. Rev. D \textbf{80}, 034011 (2009)

\bibitem{HG}
Y.~K.~Hsiao and C.~Q.~Geng,
Phys. Rev. D \textbf{91},  116007 (2015);
C.~Q.~Geng and Y.~K.~Hsiao,
  Mod.\ Phys.\ Lett.\ A {\bf 31},  1630021 (2016);
  Y.~K.~Hsiao, Y.~Yao and C.~Q.~Geng,
  Phys.\ Rev.\ D {\bf 95}, 093001 (2017).
  

\bibitem{WKL}
J.~Zhu, Z.~T.~Wei and H.~W.~Ke,
Phys. Rev. D \textbf{99},  054020 (2019)



\bibitem{MIT_bag_1} 
A.~Halprin and P.~Sorba,
Phys.\ Lett.\  {\bf 66B}, 177 (1977).


\bibitem{MIT_bag_2}
S.~Theberge and A.~W.~Thomas,
Nucl. Phys. A \textbf{393}, 252 (1983).

\bibitem{Bag48GeV}
A.~Bernotas and V.~Simonis,
Nucl. Phys. A \textbf{741}, 179 (2004).

\bibitem{MIT_bag_3}
S.~Kumar, R.~Dhir and R.~C.~Verma,
J. Phys. G \textbf{31},  141 (2005).

\bibitem{Boosting_the_bag}
M.~Betz and R.~Goldflam,
Phys. Rev. D \textbf{28}, 2848 (1983).

\bibitem{Lcwithbag}
R.~Perez-Marcial, R.~Huerta, A.~Garcia and M.~Avila-Aoki,
Phys. Rev. D \textbf{40}, 2955 (1989).

\bibitem{Isgur3}
M.~Sadzikowski and K.~Zalewski,
Z. Phys. C \textbf{59}, 677 (1993).

\bibitem{Cheng1}
H.~Y.~Cheng, X.~W.~Kang and F.~Xu,
Phys. Rev. D \textbf{97}, 074028 (2018).

\bibitem{semi}
J.~Zhu, Z.~T.~Wei and H.~W.~Ke,
Phys. Rev. D \textbf{99},  054020 (2019).


\bibitem{vectormeson}
C.~Q.~Geng, C.~W.~Liu and T.~H.~Tsai,
Phys. Rev. D \textbf{101}, 053002 (2020).


\bibitem{Cheng2}
J.~Zou, F.~Xu, G.~Meng and H.~Y.~Cheng,
Phys. Rev. D \textbf{101},  014011 (2020).

\bibitem{Korner:1994nh}
J.~G.~Korner, M.~Kramer and D.~Pirjol,
Prog. Part. Nucl. Phys. \textbf{33}, 787 (1994).


\bibitem{ABanalyze}
S.~Pakvasa, S.~P.~Rosen and S.~F.~Tuan,
Phys. Rev. D \textbf{42}, 3746 (1990).

\bibitem{Brown:1983wd}
T.~Brown, S.~F.~Tuan and S.~Pakvasa,
Phys. Rev. Lett. \textbf{51}, 1823 (1983).

\bibitem{Donoghue:1986hh}
J.~F.~Donoghue, X.~G.~He and S.~Pakvasa,
Phys. Rev. D \textbf{34}, 833 (1986).


\bibitem{ali} 
A. Ali, G. Kramer and C.D. L$\ddot{\text{u}}$, Phys. Rev.  D{\bf 58}, 094009 (1998).

\bibitem{Isgur1}
X.~Guo and P.~Kroll,
Z. Phys. C \textbf{59}, 567 (1993).

\bibitem{Isgur2}
E.~E.~Jenkins, A.~V.~Manohar and M.~B.~Wise,
Nucl. Phys. B \textbf{396}, 38 (1993).


\bibitem{Isgur4}
X.~Guo and T.~Muta,
Phys. Rev. D \textbf{54}, 4629 (1996).	
	
\bibitem{IsgurExp}
J.~Abdallah \textit{et al.} [DELPHI],
Phys. Lett. B \textbf{585}, 63 (2004).


\bibitem{Chua}
C.~K.~Chua,
Phys. Rev. D \textbf{100},  034025 (2019).
	
\bibitem{piKratio}
R.~Aaij \textit{et al.} [LHCb],
Phys. Rev. D \textbf{89},  032001 (2014).
\end{thebibliography}
\end{document}